\documentclass{article}

\usepackage{tikz, pgf, pgfplots}
\usetikzlibrary
	{
		arrows, automata, chains, datavisualization.formats.functions, fit, positioning, quantikz, quotes, shapes
	}

\usepackage{amsthm}
\usepackage{amssymb}
\usepackage[bb = boondox]{mathalfa}
\usepackage{dsfont}
\usepackage{mathtools}
\usepackage{multicol}
\usepackage{braket}
\usepackage{physics}
\numberwithin{equation}{section}

\usepackage{yquant, braket}

\usepackage[a4paper, left = 2.5cm, right = 2.5 cm, top = 2.5cm, bottom = 3.0 cm]{geometry}

\usepackage{xcolor}
\usepackage{varwidth}
\usepackage[breakable, skins, theorems]{tcolorbox}
\usepackage{graphicx}
\usepackage{hyperref}
\hypersetup
{
	colorlinks,
	linkcolor = {WordRed!75!black},
	citecolor = {WordBlueDarker25},
	urlcolor = {WordAquaDarker50}
}
\usepackage{nicematrix}
\NiceMatrixOptions
{
	code-for-first-row = \color{RedPurple},
	code-for-last-col = \color{WordAquaDarker25}
}
\usepackage{longtable}
\usepackage{colortbl}
\usepackage{float}
\usepackage{cellspace}
\usepackage{makecell}
\usepackage[boxed, ruled, vlined]{algorithm2e}
\usepackage{appendix}
\usepackage{multirow}
\newcommand{\orcidicon}[1]{\href{https://orcid.org/#1}{\includegraphics[height=\fontcharht\font`\B]{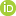}}}
\definecolor{MyLightRed}{RGB}{244, 213, 245}
\definecolor{WordRed}{RGB}{255, 0, 102}
\definecolor{RedDarkLightest}{HTML}{ff0088}
\definecolor{RedDarkLight}{HTML}{ea005f}
\definecolor{RedPurple}{HTML}{aa007f}
\definecolor{Purple}{HTML}{911146}
\definecolor{WordLightGreen}{RGB}{140, 214, 192}
\definecolor{WordGreen}{RGB}{0, 176, 80}
\definecolor{GreenLightest}{HTML}{00ffa0}
\definecolor{GreenLighter1}{HTML}{00b383}
\definecolor{GreenLighter2}{HTML}{00aa7f}
\definecolor{GreenDark}{HTML}{225522}
\definecolor{GreenTeal}{HTML}{008080}
\definecolor{WordIceBlue}{RGB}{223, 227, 229}
\definecolor{MyVeryLightBlue}{RGB}{211, 245, 247}
\definecolor{WordBlueVeryLight}{RGB}{0, 176, 240}
\definecolor{WordBlueLight}{RGB}{0, 112, 192}
\definecolor{WordBlueDark}{RGB}{46, 116, 181}
\definecolor{WordBlueDarker}{RGB}{31, 78, 121}
\definecolor{WordBlueDarker25}{RGB}{54, 96, 146}
\definecolor{WordBlueDarker50}{RGB}{36, 64, 98}
\definecolor{WordBlueDarkest}{RGB}{0, 32, 96}
\definecolor{WordBlue}{RGB}{19, 65, 99}
\definecolor{MyBlue}{RGB}{0, 64, 128}
\definecolor{MyDarkBlue}{RGB}{0, 51, 102}
\definecolor{BlueVeryDark}{HTML}{222255}
\definecolor{WordAquaLighter80}{RGB}{218, 238, 243}
\definecolor{WordAquaLighter60}{RGB}{183, 222, 232}
\definecolor{WordAquaLighter40}{RGB}{146, 205, 220}
\definecolor{WordAquaDarker25}{RGB}{49, 134, 155}
\definecolor{WordAquaDarker50}{RGB}{33, 89, 103}
\definecolor{WordVeryLightTeal}{RGB}{223, 236, 235}
\definecolor{WordLightTeal}{RGB}{160, 199, 197}
\definecolor{WordDarkTealLighter80}{RGB}{207, 223, 234}
\definecolor{WordDarkTeal}{RGB}{72, 123, 119}
\definecolor{WordDarkerTeal}{RGB}{48, 82, 80}
\definecolor{WordTurquoiseLighter80}{RGB}{209, 238, 249}
\definecolor{Brown}{HTML}{666633}

\title
	{
		Viewing biological and computer viruses
		\\
		as manifestations of games
	}

\author
	{
		Dimitris Kostadimas$^1$\orcidicon{0000-0001-6356-1946}
		and
		Kalliopi Kastampolidou$^1$\orcidicon{0000-0003-3607-9569}
		and
		Theodore Andronikos$^1$\orcidicon{0000-0002-3741-1271} \\
		$^1$Department of Informatics, Ionian University, \\
		7 Tsirigoti Square, 49100 Corfu, Greece; \\
		\{p19kost2, c17kast, andronikos\}@ionio.gr \\
	}

\begin{document}

\maketitle

\begin{abstract}
	Computer viruses exhibit many similarities with biological viruses. Thus, a closer examination of this association might lead to some new perspectives and, even, to new enhanced capabilities that will facilitate the overall effort to tackle and, why not, eradicate them. Game theory has long been considered as a useful tool for modeling viral behavior. In this paper, we establish certain, important we hope, correlations between a well-known virus, namely VirLock, with the bacteriophage $\phi6$. Moreover, following this line of thought, we also suggest efficient and, at the same time, practical strategies that may significantly alleviate the infection problems caused by VirLock and any other virus having similar traits.
\end{abstract}
\section{Introduction} \label{sec:Introduction}

\subsection{Computer viruses} \label{subsec:Computer Viruses}

The term \emph{computer virus} is an extremely common term that is used when referring to malicious computer software. This metaphorical term is based on the observation that biological viruses’ behavior and malicious software’s behavior have much in common and share some conceptual characteristics \cite{Cohen1987}.


Biological viruses have many different variations just like their digital counterparts. Computer viruses have several traits that help in their categorization, usually based on traits regarding their behavior and infection mechanisms. The way computer viruses get to infect the host computer also helps in their classification \cite{Kaspersky2021a}, \cite{uniserve}. Traits like their target demographic, their operational conditions, replication manners, infection mechanisms, infection success rate, the consequences and their severity, and many other traits are what differentiate them from each other and help in their categorization and research.

A computer virus must initially infect the target computer to interact with the host for the first time and start its replication process. In those cases where the virus implements traits like polymorphic, or metamorphic code, or a worm component, the virus’s ability to spread appears to be somewhat unpredictable.
It is not necessary for worms to manipulate the target computer software in order to replicate \cite{Norton2019}. Unless they consume a significant amount of the computer's resources, their presence will be hinted at by the slow performance. Worms remain active in the computer memory and their actions are typically unseen by the user. The most intriguing aspect of worms and viruses with worm components is not just their capacity for rapid self-replication, but also that they can do that without the host computer user’s interaction, known as the zero-click type \cite{Vick2017}, \cite{Latto2021}.



Specific types of computer viruses have the potential to mutate, which in turn leads to better spreading of their population and better infection rates while generally improving their already existing traits. Mutating is a trait found in viruses with \emph{polymorphic} and \emph{metamorphic} code. \emph{Polymorphic} code usually makes use of variable encryption to encrypt itself in order to make unique copies.
The goal of employing polymorphic and metamorphic code is to avoid detection by antimalware and antivirus technologies. Implementing the ``metamorphic trait'' in a virus is typically harder than the ``polymorphic trait.'' However, the implementation cost may be worth the extra work because it provides superior protection against antivirus technologies, and evading detection is considerably easier.


\subsection{Biological viruses} \label{subsec:Biological Viruses}

Biological viruses are parasitical organisms that gain the ability to reproduce and carry their genetic material, proteins, and DNA or RNA, by infecting a host. They translate their RNA into proteins that serve them by using the host’s ribosomes since they do not have the ability to synthesize proteins of their own. Viruses can be transmitted through different means, depending on their species. ``Host range'' is what the number of cells that get infected by a virus is called. Usually, the way that a biological virus gets dealt with is by the immune system of the organism it has infected. Infected organisms could be molecules, animals, plants, as well as humans. Moreover, a good means of defense and a way of helping the immune system is the use of vaccines, often used to work against a specific virus. Other than vaccines, there are also antiviral drugs that are evolving more and more as time passes.


When a virus infects a cell, it forces it to directly replicate itself, creating more copies of said virus. What makes up a virus is its genetic material, what is called the capsid, a set of proteins that protect said genetic material, and sometimes external lipids. The extracellular form of the virus is called the virion. Depending on whether they have a DNA or an RNA genome, the viruses are classified in two types (DNA \& RNA virus respectively). The genetic material of an RNA virus is made up of ribonucleic acid (RNA) \cite{Wagner2004}. A virus can have a lot of different effects on an organism. Causing the death of the host cell is what most of them do. Usually, they do that by using viral proteins to restrict the normal activity of the cell. The effects that some viruses have may lead to permanent damage to the host organism or they may be destroyed without malignancy. Some viruses can infect an organism without causing any changes in the cells. So, the cells can continue to function normally even when infected, yet they still end up causing the infection to spread persistently. The set of viruses that an organism gets infected by, is called a virome. A common method used to trace the source of infections is phage typing \cite{baggesen2010phage}.

Viruses do not get transmitted through cell division, since they are acellular organisms. Instead, they get transmitted by using the host to create multiple copies of themselves. The host is forced to reproduce the original virus when infected. Viruses have a basic life cycle. The infection starts once a virus attaches itself and its proteins to the surface of the host. That is when the cell type and host range get determined. After that, virions penetrate into the cell. Bacteria do not have a strong protective wall and mechanisms for gene penetration have been developed in viruses, while the capsid does not leave the outside of the cell. Eventually, the virus gets released into the host cell through the Uncoating process \cite{blaas2016viral}. When the virus is replicated the genome is also multiplied. After they replicate, altered proteins and particles may appear relative to the original form of the virus before the penetration occurred. The process that happens so that a virus can be released from a host cell is called lysis. Once this happens the cell is killed. The process the host uses to reproduce, so the virus can also be replicated, is called a prophage. Once the virus stops being inactive, lysis happens in the host cell. RNA viruses’ reproduction takes place in the cytoplasm. Each specific virus uses the enzymes it has in order to make copies of the genomes. The virus has the ability to infect a new host cell after lysis, which leads to this cycle repeating itself. Moreover, the virus might mutate during this step \cite{rogers2010bacteria}. After the immune system of an organism detects a virus, it begins the production of antibodies so that it can suppress the virus. The name of this process is humoral immunity. Whether the body has gotten rid of the virus or not, depends on the antibodies that have been produced.

Viruses that can diverse or alter microbial populations are called bacteriophages or phages and because of those properties they have been used as antibacterial agents \cite{onodera1992construction}. The host range that some bacteriophages have is focused only on one bacterial strain. Bacteriophages infect specific bacteria and are a group of viruses. More often they have double-stranded RNA genomes. RNA viruses consist of segments that make up a protein. Such segments are found in the capsid. Different segments can exist in different virions and the virus can be contagious nonetheless. They infect by attaching themselves to the molecules on the surface of the bacterium, and after that they enter the cell. Oftentimes, once the original virus enters the cell, it starts translating its mRNA into proteins. After that, the result of this process could help in the process of cell lysis or it could become a virion and help in creating other virions. Virus enzymes assist in destroying the cell membrane. Usually, bacteria use enzymes that can target unknown RNA to protect themselves from this type of infection. Bacteria also have the ability to detect the genomes of viruses that have encountered in the past and they can block their reproduction by interfering with the RNA. This is a means that bacteria use to protect themselves from this kind of infection. Bacteria can naturally interfere with the RNA. While a viral RNA is being replicated, certain mutations happen, which could either leave the cell proteins unaffected or contribute to the resistance against antiviral drugs.

\subsection{Game theory} \label{subsec:Game Theory}

There exists a tool that can, potentially, assist in modeling the behavior of such viruses and, consequently, may pave the way for the development of a higher level of protection. This tool is game theory and in its extension evolutionary game theory (EGT for short). Its relations and applications to realistic scenarios are what make this concept even more captivating \cite{weibull1997evolutionary}. Observations at the microscopic level are just as fascinating as the macroscopic ones. Even in the biological processes, game properties have been observed as well as in multicellular organisms like cells and macromolecules (see \cite{Kastampolidou2021} for an accessible overview). All of them seem to follow certain strategies, as the observation of their moves implies. Of course, during a time span, a player’s strategy can change as a result of natural selection. When mutations happen during the cell’s lifespan, reversible or irreversible changes to their strategies can take place due to epigenetic modifications. All of the aforementioned are connected to the reproduction ways of the objects at hand, and it is clear that reproductive success affects the game’s final result.

A synopsis of evolutionary games in the context of biological systems is given in \cite{Kastampolidou2020}. The use of games in biology is a significant and ongoing trend. Numerous traditional games, including the well-known Prisoner’s Dilemma, have been employed to simulate biological circumstances (see \cite{Kastampolidou2020a} and \cite{Archetti2019-ia} for references). This extends beyond viruses to include microorganisms and their games and even bio-inspired computational models (see \cite{Theocharopoulou2019} and \cite{Giannakis2015a} for further details). The study of biological processes stands to gain new perspectives and insights with the introduction of unorthodox tools like games, automata, and quantum mechanics. For instance, applying game theory concepts to the science of quantum computation has shown to be incredibly fruitful (see \cite{Giannakis2015b}, \cite{Andronikos2018}, \cite{Giannakis2019} and \cite{Andronikos2021} for some recent results and more related references). It is worth noting that games, despite their playful facade, may tackle serious, even critical problems. For instance, coin tossing plays a crucial role in the design of quantum cryptographic protocols (see \cite{Bennett2014} and references therein, and the more recent \cite{Ampatzis2021} and \cite{Ampatzis2022}).

\textbf{Contribution.} Through the correlation of computer viruses to their biological counterparts, this paper aims to offer fresh perspectives and advocate for a new, hopefully promising, vein of research. In this work we build upon the preliminary investigation of \cite{Kostadimas2021} and we demonstrate that the association of the behavioral traits of biological viruses and computer viruses is not only possible, but also fruitful. The particular focus of this research is on VirLock, a ransomware-type computer virus, and its similarities to biological viruses with an initial focus on the well-documented and studied $\phi6$ bacteriophage virus. Our approach culminates in a thorough examination and analysis of the main similarities and anticipated differences between these two viruses. In addition to improving the variety of tools for evaluating the efficacy of the strategies employed to counter the viruses, we anticipate that this approach will shed new light on the adoption and application of strategies that in the end, have proven efficient in combating certain virus types. Of course, there are numerous types of computer viruses, and the same is true for the biological ones. However, we believe that the analytical approach adopted here, can easily generalize and prove helpful in tackling more general scenarios.


\subsection{Organization} \label{subsec:Organization}

This paper is organized as follows. Section \ref{sec:Introduction} gives an introduction to the subject along with some relevant references. Sections \ref{sec:The VirLock Virus} and \ref{sec:Modeling VirLock With Game Theory} provide a concise introduction about the VirLock virus, and explain how it can be modeled using game theory, respectively. An extensive comparison of VirLock with $\phi6$ is presented in Section \ref{sec:Comparison of VirLock with $phi 6$}, while Section \ref{sec:Mathematical Formulation} is devoted to a rigorous mathematical analysis. Finally, Section \ref{sec:Discussion and Conclusions} summarizes and discusses the prospects of this approach.

\section{The VirLock virus} \label{sec:The VirLock Virus}

VirLock is a computer virus that when it manages to infect the victims computer encrypts the majority of the user's precious files while essentially locking their system. Then, demands a ransom from it's victims in order to offer them access back to their system and data. This behavior is what classifies this virus to the, now rising, ransomware group of viruses. VirLock is usually spread through cloud storage and exhibits parasitic behavior as it infects certain supported computer files. From the time it executes in the host computer, it begins to infect the supported files. It is noteworthy that the way it alters the files slightly differs from what is normally observed from similar types of malware. Instead of embedding malware within clean code, VirLock embeds clean code within malware. This implies that every encrypted file will be embedded into the malware. The file then functions as a VirLock mutation, and it can be used to further infect and spread through populations.

In 2014 the first VirLock detection took place \cite{aurangzeb2017ransomware}. Naturally, as the virus is polymorphic, numerous distinct mutations have been discovered throughout the course of many years, up until the present. Differences in VirLock's core functions as well as the \emph{decoration-code} were discovered as it continued to evolve. VirLock is capable of propagating through networks and due to the growing popularity of cloud storage nowadays, which VirLock takes advantage of, it was able to achieve better spreadability.
VirLock has the ability to take over the entire screen area of the computer and terminate the Windows {\tt explorer.exe} process, which controls the graphical user interface \cite{Sophos2016}. By the time it infects the computer, it has been rendered nearly unusable because there is no way to access the operating system's core features as the virus message covers up the entire screen while binary files and files with specific extensions are ``encrypted'' in the background. In certain VirLock variants, the user's geolocation is also breached and based on it VirLock is able to display special lock screen messages that pretend to be local authorities instead of a generic one \cite{blackberry}. This helps persuade the user even more that the message is legit and that they should cooperate. While all the above take place, the user is unable to utilize antivirus software in the traditional manner. The best technique to attempt and clean a computer from VirLock, as advised by several antivirus vendors, is to boot into the safe mode with network capabilities that Windows OS provides or use a VirLock cleaner that some specific companies provide along with manuals for the disinfection operations to get up to that point. These strategies will probably prevent VirLock from launching itself at startup. If the OS boots up properly, and the OS files are not ``encrypted,'' the user can try to disinfect the computer by running a virus scan with an antivirus or antimalware software. The type of VirLock variant that infected the host will obviously affect the likelihood of successful detection and removal of the virus, as it is quite likely that a new variant may not be recognized yet.



As previously mentioned, for known VirLock variants, certain companies provide a VirLock cleaner \cite{ESETcleaner} that claims to be able to remove the virus's leftovers and ``decrypt'' the majority (if not all) of the infected files. The user is warned that false positives might also be detected and should proceed with caution. Others even suggest a cloud access security broker that could protect the cloud storage by setting limits on certain activities and/or breaking connections when needed \cite{NetskopeBroker}, \cite{NetskopeBroker2}. VirLock’s numerous variants are what sometimes challenge antimalware software, it seems that analyzing the behavior of this type of malware would be most effective in its detection and prevention.
Live behavioral analysis capabilities in antivirus solutions provide a distinct advantage in combating such types of viruses. Mutations are not completely different from the core code as they retain certain traits. The core behavior and infection strategy, in many cases, appears to remain the same if it is not altered by a third party. It is possible that the level of protection can be increased by focusing on the way viruses react to scenarios. There are various mutations and variations of VirLock in the databases of most digital security companies. Data from the well-known website virustotal suggest that the virus's mutations and variations have different detection rates from a variety of anti-virus software. In virustotal VirLock can be seen under the names PolyRansom.B, Nabucur.A (the letter after the dot declares the variation and different ones can be seen), Win32.Cryptor and more (some of the results can be found in \cite{VirTotLock1}, \cite{VirTotLock2}, \cite{VirTotLock3}). Even though behavioral analysis from antivirus and antimalware software is considered important, there are still ways that VirLock evades their emulations by employing techniques like payload encryption and generic obscure code \cite{TheStateOfSecurity2015}, \cite{CyberHoot2020}.



The best practical approach to protect against such type of infection is to keep regular backups of at least the files that are of critical importance. Measures that could also help avoid infection and further transmission are network segmentation and keeping antivirus software up to date \cite{Orange2017}, \cite{Sjouwerman2016}. As aforementioned, another great proactive measure that could protect against VirLock and similar attacks is the use of a cloud access security broker. Though due to the complexity set-up that most of those services require it does not appear to be the best solution for maximum payoff to users with little computer knowledge. On the other hand, a known exploit of VirLock and, easy getaway to safety in case of an infection, is its disregard for Windows' volume shadow copies. Therefore, the harm can be undone if this feature of Windows OS is enabled in order to recover to a previous back-up \cite{Sophos2016}. Security specialists have also found out that VirLock has a flaw that can be exploited. By entering 64 zeros in the description key field, it is possible to trick VirLock that the ransom has been paid \cite{VirLockNJCCIC}. After that, by clicking a file, the decryption process is activated, the victim has to manually extract the original file. The drawback of this strategy is that the user will have to do this for every single file in the computer, with the risk of infecting the computer once again. The process should be done by moving the recovered files to an external drive and then formatting the one with the infected files.


Some well-known ransomware tends to use one-way encryption algorithms like RSA or AES. VirLock on the other hand does not, and instead it performs a two-stage ``encryption'' process making use of XOR and XOR-ROL operations \cite{Sophos2016}. Of course, if algorithms like RSA or AES were utilized entropy would turn out to be much higher. Even though VirLock’s mechanism is not really considered encryption, in this paper we will refer to these operations of the virus as such. The infection occurs when the user attempts to run the infectious file. By the time it is executed it drops 3 randomly named executables in random folders. Older variants of the virus appear to drop only 2 of them according to sources. The executable drops appear to carry different hashes as they are polymorphic \cite{Netskope2017}. One of them disguises itself as a Windows service, while the others encrypt and infect the computer's files. The task manager process is also disabled providing an extra layer of protection to the virus, preventing the user from taking control and killing the virus process that attempts to alter the Windows registry.



\begin{figure}[H]
	\begin{tcolorbox}
		[
			grow to left by = - 1.50 cm,
			grow to right by = - 1.50 cm,
			colback = white,					
			enhanced jigsaw,					
			sharp corners,
			toprule = 1.0 pt,
			bottomrule = 1.0 pt,
			leftrule = 0.1 pt,
			rightrule = 0.1 pt,
			sharp corners,
			center title,
			fonttitle = \bfseries
		]
		\centering
		\includegraphics[scale = 0.10, trim = {0 0 0cm 0}, clip]{"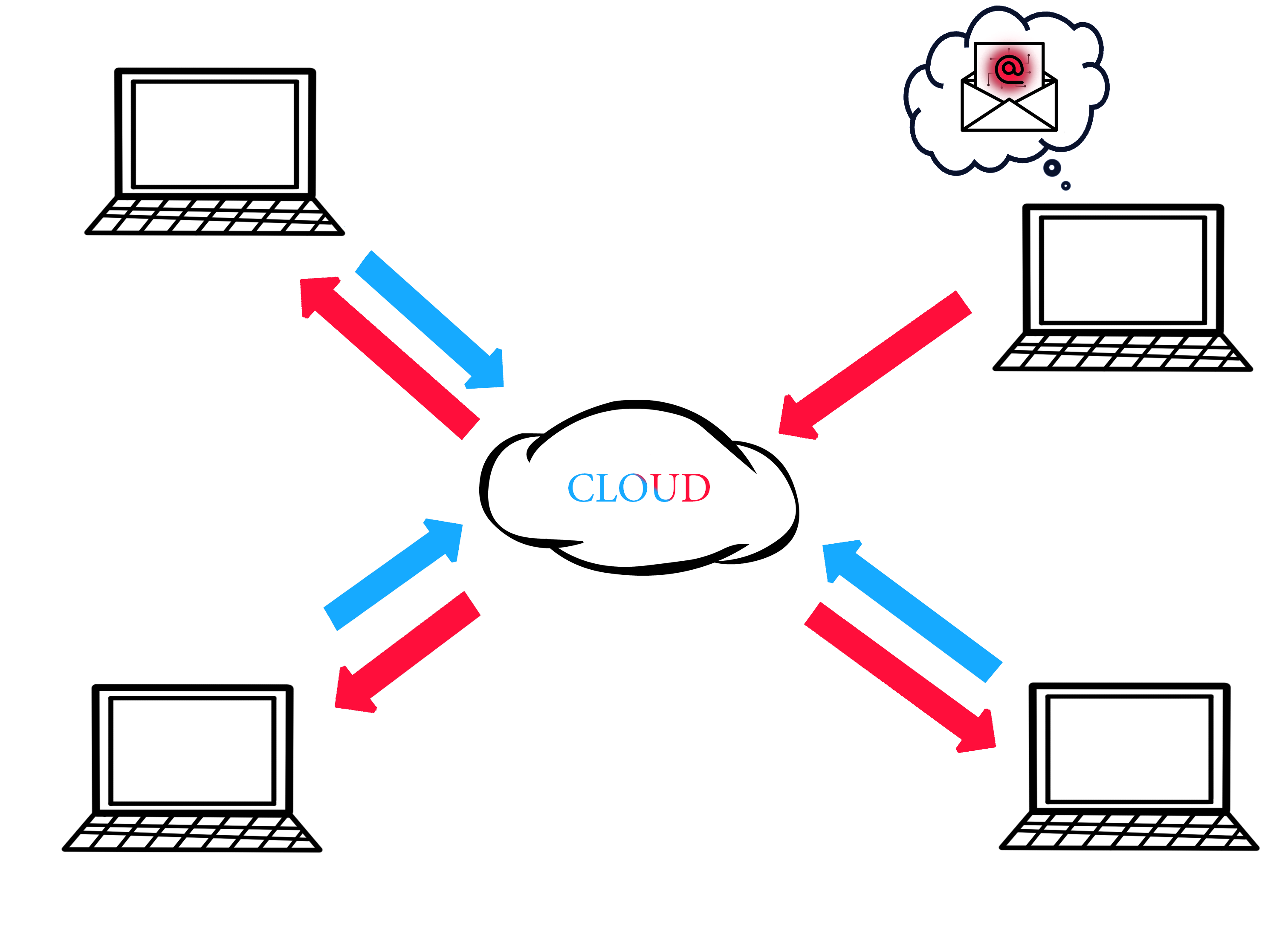"}
		\caption{This figure illustrates the VirLock cloud storage spread and infection. In the example above, the user of the top right computer opened a malicious email attachment that caused the PC to get infected with the VirLock malware. The files in the cloud storage that the computers in the network share will also get infected. The malware may eventually spread to other machines in the network, if they interact with it. The blue arrows represent interaction with the cloud, while the red ones show the route of the infection.}
		\label{fig:VirLock}
	\end{tcolorbox}
\end{figure}

The structure of these types of files is the following: polymorphic code appears at the begging as well as the end of the code that changes in every iteration/infection. Polymorphic code is essentially ``wrapped'' around the main code blocks. This part of the code is also referred to as ``decoration code'' as it ``decorates'' the main code with operations like random API calls from random modules. The malicious code, which runs every time, appears after the first piece of polymorphic code and it is usually the same among different variations of the code. Right after the malicious code, VirLock embeds the encrypted data of the original file it infected, often referred to as ``clean code.'' The last piece of polymorphic code appears at the end of the code \cite{Sjouwerman2016}, \cite{knowbe4New}, \cite{vbc2015Vir}. In case of an infection, it is possible that the user might be tempted to pay the ransom rather than getting involved in disinfection operations that might appear complicated. There are several major reasons why one should reconsider making the ransom payment. First and foremost, there is a compelling ethical justification, especially in light of the exorbitant price the ransomware may demand. Second, according to a number of sources \cite{Sophos2021}, just 8\% of those who pay the ransom successfully recover the grand total of their data. Ransomware counts on the desperation of its victims to get their data back, pushing them to the ransom payment strategy. Of course, as it appears, there are other significantly more efficient ways for victims to get their data back, while achieving a high pay-off as players.

\section{Modeling VirLock with game theory} \label{sec:Modeling VirLock With Game Theory}


It is gainful to acquire a clearer picture of the strategies that benefit the user the most in case of an infection from VirLock by employing game theory tools to model various scenarios. Based on the general consensus among security experts and information from \cite{Sophos2021} that ``96\% of those whose data was encrypted got their data back in the most significant ransomware attack,'' and that ``only 8\% (of those who paid) got all their data back'' the payoff matrix that follows can be constructed. This confirms that there are other, more effective methods of recovering the data besides paying the ransom. The following game theory matrix pictures a scenario where the a user who got infected by ransomware has to decide whether paying the ransom or not is worth it.

\begin{figure}[H]
	\begin{tcolorbox}
		[
			grow to left by = - 1.50 cm,
			grow to right by = - 1.50 cm,
			colback = white,					
			enhanced jigsaw,					
			sharp corners,
			toprule = 1.0 pt,
			bottomrule = 1.0 pt,
			leftrule = 0.1 pt,
			rightrule = 0.1 pt,
			sharp corners,
			center title,
			fonttitle = \bfseries
		]
		\centering
		\includegraphics[scale = 0.10, trim = {0 0 0cm 0}, clip]{"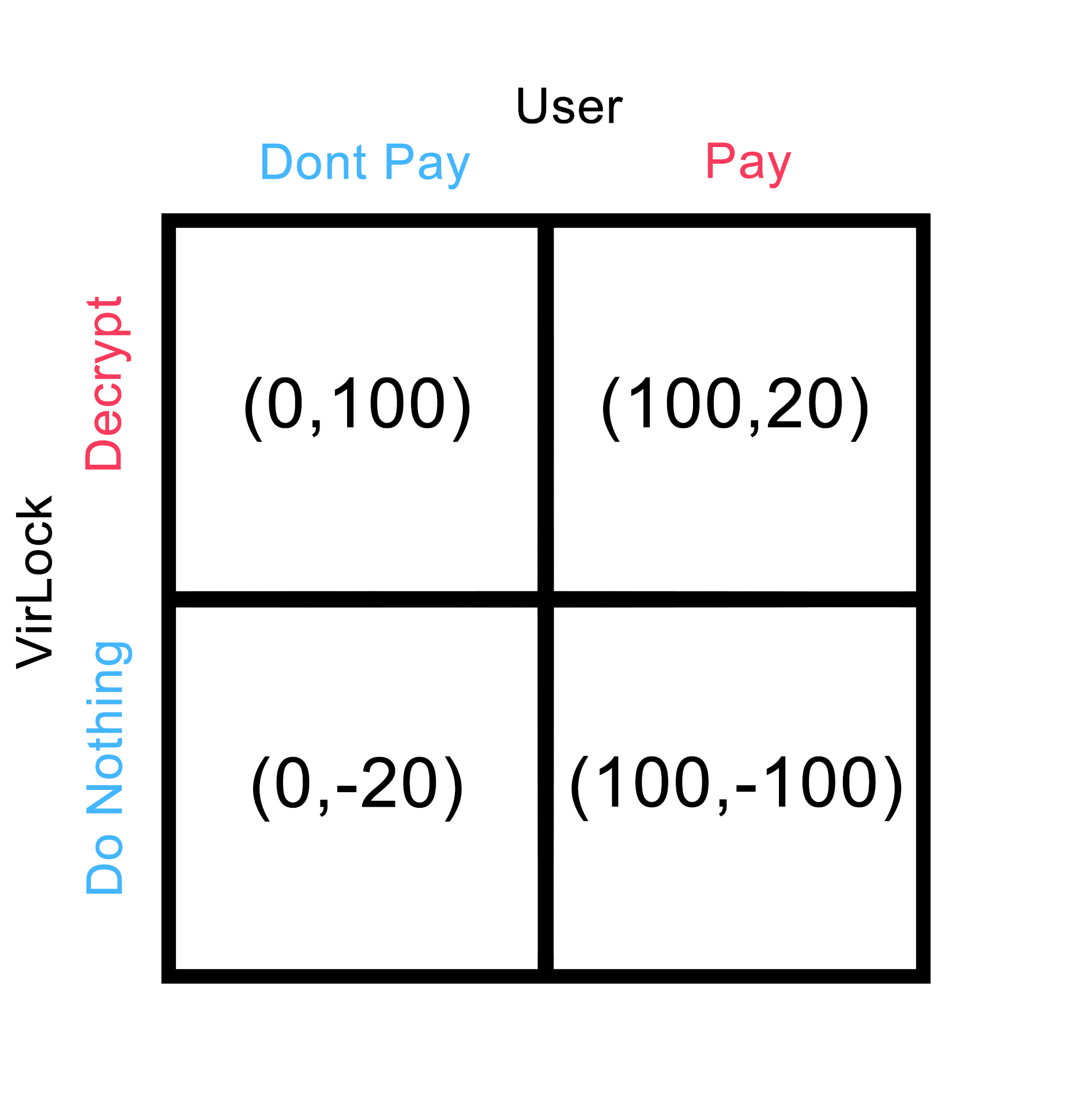"}
		\caption{Game theory payoff matrix of the ransom payment. The VirLock malware may or may not decrypt the users' data (or at least not in its entirety), and on the other hand, the user has the choice of paying or refusing to pay the ransom. The above payoff matrix makes it apparent that paying the ransom payment will generally result in advantageous position for the malware's creator(s) and that doing so entails extra risks.}
		\label{fig:Payoff Matrix}
	\end{tcolorbox}
\end{figure}

The game pictured above may appear to be simple, but it's a helpful introduction to the concept. In this game the VirLock malware as well as the user of the computer are considered to be the players. The user has two possible moves and so does VirLock. The user can choose to pay or not to pay the ransom and VirLock can ``choose'' to decrypt or not to decrypt the files. VirLocks' moves are technically predetermined since usually there is an algorithm that offers the user the decryption key but we can still roughly calculate the possible outcomes of the scenario. While keeping in mind the aforementioned statistics we have improvised on the payoff outcome of every strategy combination. The max payoff is considered to be $100$. For the user, $100$ means that all his files were decrypted without losing any money (which does not appear here since we don't consider certain disinfection techniques but only ransom payment). For VirLock $100$ means that the entirety of the ransom was paid. We consider that the data is at least of some importance.

Each table cell represents the outcome of the game after the players have picked their corresponding strategies. On the top left cell we have the statistically improbable case where the files are decrypted while the user doesn't pay. This case can be considered as a failure of VirLock to encrypt the data or the user employed a proactive strategy so no damage was taken from the attack. Because of that the payoff for the user was taken to be $100$ since not only no money was spent but all the data were unaffected while tackling the attack and preventing further spread. On the other hand, in the bottom left cell we have the case where the user doesn't pay and VirLock doesn't decrypt. This could be considered a neutral state but in reality if VirLock managed to already encrypt the data the user is in a disadvantage. Of course, based on the criticality of the data and the possible disinfection techniques per case the negative number could go even lower, meaning bigger loses for the user. We will further demonstrate the effects of such factors in the following sections. 

On the top right block, the user chooses to pay the ransom and VirLock decrypts the files. As the available statistics suggest, only a portion of the data is usually getting decrypted. Moreover, considering that the ransom is quite high, the user benefits only in theory, since usually not all files are rescued, and only some of the data are recovered, despite the significant amount paid.
Again, considering the importance of the files the payoff is proportionate to the cost of money and the criticality of the data that are now encrypted. The benefit could even go bellow $0$ depending on the circumstances. The last bottom left cell represents the case where the users pays the whole ransom and VirLock does nothing to decrypt the data or a second infection takes place from user error or VirLocks' inability. In this case the users loses everything in the context of this game with cost $-100$ while VirLock gets it all ($100$). Analyzing certain scenarios that way, can provide a better insight and understanding of the actual costs and the benefits of strategy combinations between players.

\section{Comparison of VirLock with $\phi6$} \label{sec:Comparison of VirLock with $phi 6$}

It is of considerable interest, and may also yield new insights, to compare the properties of biological and computer viruses and how they behave during and after the invasion of the host. Computer virus types continue to evolve in tandem with the development of computers, just like their biological counterparts. Biological viruses mutate and when they do, generations can be observed over time. In a similar fashion, computer viruses may also mutate and replicate themselves or can be altered by a third party. To follow this metaphor even further, we may correlate the immune system of a bacterium to the antivirus software and other security software installed on a computer as well as other digital security properties (like privileges, actions, settings et.c.). By infecting the victim's computer files (which in this case could represent the cells of a bacterium that get infected from a biological virus), computer viruses attempt to weaken the computer's ``immune system'' in order to replicate themselves and increase their population. In addition to the above, VirLock and worm-type viruses try to infect more than just a single individual and spread their population through the network of computers, which in turn can be correlated to a population of bacteria in close proximity. As we attempt to compare and correlate computer viruses with their biological counterparts we decided on a representative of each group. We deemed the bacteriophage $\phi 6$ as a great representative of the biological viruses since it has been extensively modeled in terms of its behavior and structure, and has previously been associated with the field of classical and evolutionary game theory.

\subsection{The pseudomonas virus $\phi6$} \label{subsec:The Pseudomonas Virus φ6}

Bacteriophage $\phi6$ lytic virus is a member of the cystoviridae family infecting Pseudomonas bacteria. Specifically, that type of virus infects legume-infecting bacteria. It's double-stranded genome consists of twelve protein codes and three segmented parts. A lipid membrane covers the nucleocapsid of such species. With its unique protein, named P$3$, designed for this function, $\phi6$ finds the bacterium that it wishes to infect and then sticks to it. In addition to the protein above, many other proteins participate in the process of cell infection. $\phi6$ bacteriophages compete with each other to gain capsid proteins in their common bacterial targeted host. When different strains of phi-6 bacteriophages are assembled, with varying numbers of bacteriophage particles that can simultaneously infect a single bacterial cell, different results are obtained. The result of infection of the bacterial cell by the virus directly depends on the combination and proportions that will be chosen in the populations of phage and bacteria. A strain can infect a cell alone, or infection of the cell can result from a combination of viruses and their counterparts or even their mutations \cite{turner2005cheating}.

The behavior of the bacteriophages, as well as the role that each one will take and whether they will cooperate with their neighbors or show selfish behavior plays a catalytic role in the fitness of the population against the virus. $\phi6$ is found in cases of polymorphism mix with helper viruses, for the purpose of ESS of the game. Some behaviors are interpreted with the help of the prisoner's dilemma and others with the help of the snowdrift game (also called the chicken game) with different results each, depending on the ratio of infecting viruses to bacterial cells \cite{turner2003escape}.

The main properties and characteristics of VirLock that could be linked to bacteriophages and especially $\phi6$ are the following:

\begin{tcolorbox}
	[
		grow to left by = 0.00 cm,
		grow to right by = 0.00 cm,
		colback = GreenLighter1!03,			
		enhanced jigsaw,					
		sharp corners,
		toprule = 1.0 pt,
		bottomrule = 1.0 pt,
		leftrule = 0.1 pt,
		rightrule = 0.1 pt,
		sharp corners,
		center title,
		fonttitle = \bfseries
	]
	\renewcommand\labelenumi{(\theenumi)}
	\begin{enumerate}
		\item	They are self-replicating viruses that appear to grow rapidly.
		\item	They try to protect themselves by attacking and eventually manipulating the host.
		\item	They affect the host in order to gain full access to its functions and keep their viral ability unaffected.
		\item	They affect certain host types.
		\item	They exhibit parasitic behavior, while manipulating and embedding their code in their replicants/mutants.
		\item	They are able to spread when the infected parts come in contact with other hosts.
		\item	They have a core structure that is subject to change.
		\item	They are able to mutate rapidly not only by themselves but also with external help.
		\item	They can be untraceable for a specific time; they are, in general, hard to locate and extremely difficult to eliminate.
	\end{enumerate}
\end{tcolorbox}

Table \ref{tbl:General Correlation Hypothesis} summarizes the main thesis of our approach regarding the fundamental correlation between computer and biological viruses. Tables \ref{tbl:Similarities between VirLock and $phi 6$} and \ref{tbl:Differences between VirLock and $phi 6$} elaborate on the main similarities and differences between VirLock and $\phi6$, respectively.

\begin{table}[H]
	\renewcommand{\arraystretch}{1.30}
	\begin{tcolorbox}
		[
			grow to left by = - 0.25 cm,
			grow to right by = - 0.25 cm,
			colback = gray!03,
			enhanced jigsaw,					
			sharp corners,
			boxrule = 0.1 pt,
			toprule = 0.1 pt,
			bottomrule = 0.1 pt
		]
		\caption{This table presents the hypothesis regarding the correlation between computer and biological viruses in general.\\}
		\label{tbl:General Correlation Hypothesis}
		\centering
		\begin{tabular}
			{
				>{\centering\arraybackslash} m{6.00 cm} !{\vrule width 1.25 pt}
				>{\centering\arraybackslash} m{6.00 cm} 
			}
			\Xhline{4\arrayrulewidth}
			\multicolumn{2}{c}{\textbf{General correlation hypothesis}}
			\\
			\Xhline{\arrayrulewidth}
			Computer networks
			&
			Cities/Countries
			\\
			\Xhline{3\arrayrulewidth}
			PC in a computer network
			&
			Individual in a city or country
			\\
			\Xhline{1\arrayrulewidth}
			Infected files in a computer system
			&
			Infected cells in a human organism
			\\
            \Xhline{1\arrayrulewidth}
			Security software and properties (e.g. antivirus, privileges, actions)
			&
			Immune system of a human individual
			\\
			\Xhline{4\arrayrulewidth}
		\end{tabular}
	\end{tcolorbox}
	\renewcommand{\arraystretch}{1.0}
\end{table}

\begin{table}[H]
	\renewcommand{\arraystretch}{1.30}
	\begin{tcolorbox}
		[
			grow to left by = - 1.00 cm,
			grow to right by = - 1.00 cm,
			colback = gray!03,
			enhanced jigsaw,					
			sharp corners,
			boxrule = 0.1 pt,
			toprule = 0.1 pt,
			bottomrule = 0.1 pt
		]
		\caption{This table highlights the similarities between VirLock and $\phi6$ \\}
		\label{tbl:Similarities between VirLock and $phi 6$}
		\centering
		\begin{tabular}
			{
				>{\centering\arraybackslash} m{6.00 cm} !{\vrule width 1.25 pt}
				>{\centering\arraybackslash} m{6.00 cm} 
			}
			\Xhline{4\arrayrulewidth}
			\multicolumn{2}{c}{\textbf{Similarities}}
			\\
			\Xhline{\arrayrulewidth}
			VirLock
			&
			$\phi6$
			\\
			\Xhline{3\arrayrulewidth}
			Infects certain file extensions (.xls, .doc, .pdf, .rtf, .psd, .dwg, .cdr, .cd, .mdb, .lcd, .dbf, .sqlite, .jpg, .zip)
			&
			Pseudomonas bacteria
			\\
			\Xhline{1\arrayrulewidth}
			Parasitic
			&
			Parasitic
			\\
			\Xhline{1\arrayrulewidth}
			Polymorphic code, mutations in iterations
			&
			Polymorphism/mutations to virus \& host cells
			\\
			\Xhline{1\arrayrulewidth}
			Replicates after infection
			&
			Replicates after infection
			\\
			\Xhline{1\arrayrulewidth}
			Embeds clean code inside malware
			&
			Alters the RNA of host ribosomes into proteins that serve them
			\\
			\Xhline{1\arrayrulewidth}
			Antivirus \& antimalware
			&
			Immune system, RNA interference
			\\
			\Xhline{1\arrayrulewidth}
			VirLock cleaner
			&
			Vaccines \& antiviral drugs
			\\
			\Xhline{1\arrayrulewidth}
			Permanent damage to files (damage to the computer is avoidable)
			&
			Permanent damage to host cells
			\\
			\Xhline{1\arrayrulewidth}
			Infected files can also infect
			&
			Infected cells can also infect
			\\
			\Xhline{4\arrayrulewidth}
		\end{tabular}
	\end{tcolorbox}
	\renewcommand{\arraystretch}{1.0}
\end{table}

\begin{table}[H]
	\renewcommand{\arraystretch}{1.30}
	\begin{tcolorbox}
		[
			grow to left by = - 1.00 cm,
			grow to right by = - 1.00 cm,
			colback = gray!03,
			enhanced jigsaw,					
			sharp corners,
			boxrule = 0.1 pt,
			toprule = 0.1 pt,
			bottomrule = 0.1 pt
		]
		\caption{This table highlights the differences between VirLock and $\phi6$.\\}
		\label{tbl:Differences between VirLock and $phi 6$}
		\centering
		\begin{tabular}
			{
				>{\centering\arraybackslash} m{6.00 cm} !{\vrule width 1.25 pt}
				>{\centering\arraybackslash} m{6.00 cm} 
			}
			\Xhline{4\arrayrulewidth}
			\multicolumn{2}{c}{\textbf{Differences}}
			\\
			\Xhline{\arrayrulewidth}
			VirLock
			&
			$\phi6$
			\\
			\Xhline{3\arrayrulewidth}
			E-mail, transfer infected file, cloud
			&
			Virus enzymes
			\\
			\Xhline{1\arrayrulewidth}
			Multiple attributes
			&
			P12 proteins with different attributes
			\\
			\Xhline{1\arrayrulewidth}
			Polymorphic code, malware code, clean code, polymorphic
			&
			RNA, Capsid, Virion
			\\
			\Xhline{4\arrayrulewidth}
		\end{tabular}
	\end{tcolorbox}
	\renewcommand{\arraystretch}{1.0}
\end{table}

The following Tables \ref{tbl:Complexity of VirLock Recovery Strategies}--\ref{tbl:Antivirus + Cleaner} describe the steps of some of the most prominent strategies that users can follow in order to recover their computer back to a normal functioning state. The rationale behind the following tables is to get an overall sense of the complexity inherent in each step, as well as the effectiveness and the risks that it entails in order to get a better insight of the overall costs and benefits that could be used in this type of game.

\begin{table}[H]
	\renewcommand{\arraystretch}{1.30}
	\begin{tcolorbox}
		[
			grow to left by = 0.50 cm,
			grow to right by = 0.50 cm,
			colback = gray!03,
			enhanced jigsaw,					
			sharp corners,
			boxrule = 0.1 pt,
			toprule = 0.1 pt,
			bottomrule = 0.1 pt
		]
		\caption{This table shows the complexity of VirLock recovery strategies.\\}
		\label{tbl:Complexity of VirLock Recovery Strategies}
		\centering
		\begin{tabular}
			{
				>{\centering\arraybackslash} m{6.00 cm} !{\vrule width 1.25 pt}
				>{\centering\arraybackslash} m{2.00 cm} !{\vrule width 1.25 pt}
				>{\centering\arraybackslash} m{2.50 cm} !{\vrule width 1.25 pt}
				>{\centering\arraybackslash} m{3.00 cm} 
			}
			\Xhline{4\arrayrulewidth}
			\multicolumn{4}{c}{\textbf{Recovery strategy complexity}}
			\\
			\Xhline{\arrayrulewidth}
			Strategy
			&
			Complexity (out of $10$)
			&
			Effectiveness
			&
			Risk of reinfection
			\\
			\Xhline{3\arrayrulewidth}
			Ransom payment
			&
			1
			&
			Low
			&
			High
			\\
			\Xhline{1\arrayrulewidth}
			Decrypt taking advantage of VirLock's flaw
			&
			5
			&
			Medium
			&
			High
			\\
			\Xhline{1\arrayrulewidth}
			Recovery using shadow volume copies
			&
			4
			&
			High (depends)
			&
			Medium
			\\
			\Xhline{1\arrayrulewidth}
			Simple malware removal with antivirus software
			&
			6
			&
			High
			&
			Low
			\\
			\Xhline{1\arrayrulewidth}
			Virus removal and special cleaner (antivirus + cleaner)
			&
			8
			&
			High
			&
			Low
			\\
			\Xhline{4\arrayrulewidth}
		\end{tabular}
	\end{tcolorbox}
	\renewcommand{\arraystretch}{1.0}
\end{table}

\begin{table}[H]
	\renewcommand{\arraystretch}{1.30}
	\begin{tcolorbox}
		[
			grow to left by = - 0.25 cm,
			grow to right by = - 0.25 cm,
			colback = gray!03,
			enhanced jigsaw,					
			sharp corners,
			boxrule = 0.1 pt,
			toprule = 0.1 pt,
			bottomrule = 0.1 pt
		]
		\caption{The recovery strategy of decrypting taking advantage of VirLock's flaw.\\}
		\label{tbl:Decrypt Taking Advantage of VirLock's Flaw}
		\centering
		\begin{tabular}
			{
				>{\centering\arraybackslash} m{6.00 cm} !{\vrule width 1.25 pt}
				>{\centering\arraybackslash} m{6.00 cm} 
			}
			\Xhline{4\arrayrulewidth}
			\multicolumn{2}{c}{\textbf{Decrypt taking advantage of VirLock's exploit}}
			\\
			\Xhline{\arrayrulewidth}
			Steps
			&
			Complexity (out of $10$)
			\\
			\Xhline{3\arrayrulewidth}
			Enter $64$ zeros in the decryption key field
			&
			1
			\\
			\Xhline{1\arrayrulewidth}
			Click in every file of the computer
			&
			8
			\\
			&
			(depends since it is more time consuming than complex)
			\\
			\Xhline{4\arrayrulewidth}
		\end{tabular}
	\end{tcolorbox}
	\renewcommand{\arraystretch}{1.0}
\end{table}

\begin{table}[H]
	\renewcommand{\arraystretch}{1.30}
	\begin{tcolorbox}
		[
			grow to left by = - 0.25 cm,
			grow to right by = - 0.25 cm,
			colback = gray!03,
			enhanced jigsaw,					
			sharp corners,
			boxrule = 0.1 pt,
			toprule = 0.1 pt,
			bottomrule = 0.1 pt
		]
		\caption{The recovery strategy of using shadow volume copies.\\}
		\label{tbl:Using Shadow Volume Copies}
		\centering
		\begin{tabular}
			{
				>{\centering\arraybackslash} m{6.50 cm} !{\vrule width 1.25 pt}
				>{\centering\arraybackslash} m{6.00 cm} 
			}
			\Xhline{4\arrayrulewidth}
			\multicolumn{2}{c}{\textbf{Recover using shadow volume copies}}
			\\
			\Xhline{\arrayrulewidth}
			Steps
			&
			Complexity (out of $10$)
			\\
			\Xhline{3\arrayrulewidth}
			Have shadow volume copies enabled and available beforehand
			&
			2
			\\
			\Xhline{1\arrayrulewidth}
			Boot into the Windows OS in safe mode
			&
			4
			\\
			\Xhline{1\arrayrulewidth}
			Recover to a previous shadow copy
			&
			4
			\\
			\Xhline{4\arrayrulewidth}
		\end{tabular}
	\end{tcolorbox}
	\renewcommand{\arraystretch}{1.0}
\end{table}

\begin{table}[H]
	\renewcommand{\arraystretch}{1.30}
	\begin{tcolorbox}
		[
			grow to left by = - 0.25 cm,
			grow to right by = - 0.25 cm,
			colback = gray!03,
			enhanced jigsaw,					
			sharp corners,
			boxrule = 0.1 pt,
			toprule = 0.1 pt,
			bottomrule = 0.1 pt
		]
		\caption{The recovery strategy of using simple malware removal with antivirus.\\}
		\label{tbl:Malware Removal with Antivirus}
		\centering
		\begin{tabular}
			{
				>{\centering\arraybackslash} m{6.50 cm} !{\vrule width 1.25 pt}
				>{\centering\arraybackslash} m{6.00 cm} 
			}
			\Xhline{4\arrayrulewidth}
			\multicolumn{2}{c}{\textbf{Malware removal with antivirus}}
			\\
			\Xhline{\arrayrulewidth}
			Steps
			&
			Complexity (out of $10$)
			\\
			\Xhline{3\arrayrulewidth}
			Boot into the Windows OS in safe mode
			&
			4
			\\
			\Xhline{1\arrayrulewidth}
			Install an antivirus using an external device
			&
			4 (not always necessary)
			\\
			\Xhline{1\arrayrulewidth}
			Scan the device for malware
			&
			2
			\\
			\Xhline{4\arrayrulewidth}
		\end{tabular}
	\end{tcolorbox}
	\renewcommand{\arraystretch}{1.0}
\end{table}

\begin{table}[H]
	\renewcommand{\arraystretch}{1.30}
	\begin{tcolorbox}
		[
			grow to left by = 0.25 cm,
			grow to right by = 0.25 cm,
			colback = gray!03,
			enhanced jigsaw,					
			sharp corners,
			boxrule = 0.1 pt,
			toprule = 0.1 pt,
			bottomrule = 0.1 pt
		]
		\caption{The recovery strategy of using virus removal \& cleaner with recovery features (antivirus + cleaner).\\}
		\label{tbl:Antivirus + Cleaner}
		\centering
		\begin{tabular}
			{
				>{\centering\arraybackslash} m{7.75 cm} !{\vrule width 1.25 pt}
				>{\centering\arraybackslash} m{6.00 cm} 
			}
			\Xhline{4\arrayrulewidth}
			\multicolumn{2}{c}{\textbf{Recover using antivirus + cleaner}}
			\\
			\Xhline{\arrayrulewidth}
			Steps
			&
			Complexity (out of $10$)
			\\
			\Xhline{3\arrayrulewidth}
			Boot into the Windows OS in safe mode
			&
			4
			\\
			\Xhline{1\arrayrulewidth}
			Install an antivirus using an external device
			&
			4 (not always necessary)
			\\
			\Xhline{1\arrayrulewidth}
			Install a VirLock cleaner using an external device
			&
			4 (not always necessary)
			\\
			\Xhline{1\arrayrulewidth}
			Run the cleaner (requires several steps and might result in deleting files that are not infected
			&
			5
			\\
			\Xhline{1\arrayrulewidth}
			Scan the device for malware
			&
			2
			\\
			\Xhline{4\arrayrulewidth}
		\end{tabular}
	\end{tcolorbox}
	\renewcommand{\arraystretch}{1.0}
\end{table}

We rated the above steps based on how difficult it would be for a typical computer user to carry out one of these operations. The \emph{complexity} variable has a scale from $0$ to $10$. $0$ meaning that any average user could complete this task without any issues, whereas $10$ means that the steps are either time consuming, dangerous, too complex (requiring special knowledge and have sub-steps) or there are factors that have other requirements. Depending on the relative success rate of a certain technique and the percentage of recovered files (in the event that the recovery of all of them is not possible), the \emph{effectiveness} variable can have one of three possible values: low, medium, or high. When employing a specific recovery method, the user might be susceptible to contracting the virus once more. To quantify the risk in such an eventuality, the \emph{re-infection} variable can be used as an indicator of whether the user is more or less susceptible to re-infection and uses the same scale as the \emph{effectiveness} variable.

The previous tables (as well as the ransom payment payoff matrix of Figure \ref{fig:Payoff Matrix}) strive to assist the unlucky users that VirLock managed to infect their computer in order to choose a strategy that may offer them the highest pay-off against VirLock in their unique infection scenario. Additionally, individuals that have been infected by similar software may find the tables informative. The user can clearly benefit from having a well-thought-out plan in place beforehand (like those tables and game theory, in general, can provide). As aforementioned this can also help to analyze costs and benefits.

The recovery technique utilizing shadow volume copies has indeed a high success rate. However, whether it is possible at all or not, depends on how proactive the user is in terms of keeping shadow volume copies beforehand and how old they are. Moreover, depending on how many files the user stores in the computer, the ``click on every file in your computer'' step may not be a great option, but also it might not be needed at all depending on the situation. We point out that installation of an antivirus or antimalware software is not always necessary after the infection has taken place if they have been already installed beforehand and are still operational. The strategies above are broken into steps, and the difficulty value associated to each step is chosen with the average computer user in mind. Of course these values are subject to change based on the traits of the individual computer user. We should disclaim that, obviously, a computer science graduate or a high prestige company employee would make the complexity variable have much lower values, but this is not always the case.

\section{Mathematical formulation} \label{sec:Mathematical Formulation}

\subsection{Intuition and derivation} \label{subsec:Intuition & Derivation}

By taking into account all the aforementioned facts and measurements and by adding to the picture real-life factors, like the common behavior of a user during a computer virus infection scenario, as well as known factual data, we end up with conceptual mathematical equations and formulas. The underlying mathematical formulation not only supports procedures like the simple calculation of the severity of a virus, but can also help us factor out the importance of those realistic factors.

A great example that is conceptually close is set by the Common Vulnerability Scoring System (CVSS) which is a free and open industry standard, that identifies the key technical aspects of software, hardware, and firmware vulnerabilities. It generates numerical scores that indicate the severity of a vulnerability in comparison to other vulnerabilities. The scores are based on realistic characteristics of viruses and vulnerabilities, like the scope of the vulnerability, the user interaction, the availability, the level of possible remediation and many more. All the above metrics take similar values to the ones we used in the tables previously introduced. Finally, utilizing all the above, CVSS uses special formulas (like those we will present later on) to calculate a severity score that ranges from 0 to 10, with 10 being the critical severity level. Further info about CVSS can be found in this specification document \cite{FirstCVSS}, and a CVSS calculator is also provided in \cite{FirstCVSSCalc}. Even though CVSS ratings for the severity levels are approximate, the system allows ``players'' to prepare more and get the necessary resources (for example a cloud access security broker or an antivirus) to achieve maximum payoff against the known threats \cite{WikiCVSS}.

The existence of precise mathematical formulas enables further tests and experiments that may lead us to a definitive conclusion regarding the importance of the realistic factors during the infection (apart from the importance of the technical and technological ones). Games that make use of game theory could be created using these formulas in order to simulate certain infection scenarios. Then, by observing the players and the outcomes of the employed strategies, new strategies may be devised that are sure to offer advantages over the other existing ones (if any). We would like to create formulas that take into account the different types of end users and are specific for certain scenarios, so they can be more accurate than formulas utilizing the one size fits all approach. Again, in the field of biological viruses, there are several cases where formulas that factor realistic data are being used. An obvious case that comes to mind is the recent Covid-19 virus pandemic out-brake, where certain equations have been proposed \cite{RoyalACov}, \cite{ElseCov}. Nonetheless, it would seem that such formulas (that constitute a precaution measure) are not being used widely in the field of computer virus attacks.

At this point, taking into account some of the factors that are extensively mentioned in the literature, and which can affect the behavior and the outcome of a possible infection, formulas that fit certain scenarios can be designed. The following example is based on a hypothetical ransomware attack (identical or similar to VirLock). For demonstration purposes we could also correlate a small city and its population getting infected by a biological virus to a computer network getting infected by a computer virus. Inspired from the CVSS and other formulas from the computer and the biological world, as well as the typical model of a commercial company, user profile and their characteristic behavior, we propose a group of novel formulas, which, combined with the standard game theory mechanics, we believe they can prove useful in the continuous endeavor to discover better strategies to cope with viral infection. Below we list the factors we have deemed particularly important to be taken into account in well-thought-out formulas of practical value, based on a VirLock, or an even more general, ransomware attack.

\vspace{0.3 cm}

\begin{tcolorbox}
	[
		grow to left by = 0.00 cm,
		grow to right by = 0.00 cm,
		colback = RedPurple!03,				
		enhanced jigsaw,					
		sharp corners,
		toprule = 1.0 pt,
		bottomrule = 1.0 pt,
		leftrule = 0.1 pt,
		rightrule = 0.1 pt,
		sharp corners,
		center title,
		fonttitle = \bfseries
	]
	\begin{itemize}
		\item $A$ : User’s awareness of the computer virus / User's computer literacy-knowledge / Sudden anxiety from the attack
		\item $B$ : Economical state of the user / company
		\item $C$ : Criticality of the encrypted data / Amount of the critical encrypted data
		\item $D$ : Amount of data
		\item $E$ : Amount of data the virus infects
		\item $F$ : Percentage of infected computers in the network
		\item $G$ : Known ways of effective disinfection / Possible ways of data recovery
		\item $H$ : The effectiveness of the known disinfection strategies all together (Percentage based on users that attempted the strategies)
		\item $I$ : Safety of operations during/after the infection
	\end{itemize}
\end{tcolorbox}

The following formula is used to estimate the \emph{spreadability score}, denoted by $SPS$:

\begin{align} \label{eq:Spreadability Score}
	SPS = 0.7 (100 - A) + 0.3 F
	\ .
\end{align}

Given the spreadability score, the \emph{severity} of the infection, denoted by $S$, can be computed by the next formula, in which parameter $G$ is assumed to be $> 0$.

\begin{align} \label{eq:Severity}
	S = 0.1 C + 0.25 E + 0.1 F + 0.25 \, SPS + 0.3 G
\end{align}

In an analogous manner, the \emph{disinfection probability} $DP$ is given by

\begin{align} \label{eq:Disinfection Probability}
	DP = 0.15 A + 0.2 B + 0.1 (100 - E) + 0.15 (100 - F) + 0.3 H + 0.1 I
	\ .
\end{align}

Finally, the disinfection payoff, denoted by $DC$, can be computed by the Algorithm \ref{alg: Disinfection Caliber-Worthiness}, given below.

\begin{algorithm}	
	\caption{ Disinfection payoff }
	\label{alg: Disinfection Caliber-Worthiness}
	\If { $( C \leq 0.2 \ \mathrm{OR} \ S < 0.2 )$ }
	{
		$DC = 0$
	}
	\ElseIf { $(C > 0.8)$ }
	{
		$DC = C$
	}
	\ElseIf { $(S \leq 0.8)$ }
	{
		$DC = C * S$
	}
	\ElseIf { $(S \leq 1)$ }
	{
		$DC = C$
	}
	\Return $100 * DC$
\end{algorithm}

The above formulas almost always return a decimal number which represents the corresponding score in the scale of $1 - 100$ (percentage). The values of the variables range from $1$ to $100$ and determine the final output probability. For example, if only $1/4$ of the computers in the network got infected the value of the $F$ variable would be $25$. If the encrypted data a company or an individual have are of an extreme critical importance and loss would mean huge drawbacks, then high criticality could range from $90$ to $100$ for the variable $C$. Having a certain percentage scale would help having more accurate numbers in pay-off matrices making sure that there's a certain amount of benefits both players can use in their advantage. Of course the above have certain flaws and are made up just for demonstration.

\subsection{Test case scenario} \label{subsec:Test Case Scenario}

To get a better understanding of previous mathematical concepts, we may visualize them with the following graphical representations. At the same time, this will allow us to compile a general profile of the players, which in this scenario are $2$ companies with different traits against VirLock.

\begin{table}[H]
	\renewcommand{\arraystretch}{1.30}
	\begin{tcolorbox}
		[
			grow to left by = - 2.00 cm,
			grow to right by = - 2.00 cm,
			colback = gray!03,
			enhanced jigsaw,					
			sharp corners,
			boxrule = 0.1 pt,
			toprule = 0.1 pt,
			bottomrule = 0.1 pt
		]
		\caption{The values of the variables used to model the profiles of the fictional companies A and B used in the example scenarios that follow.\\}
		\label{tbl:Variables & Values}
		\centering
		\begin{tabular}
			{
				>{\centering\arraybackslash} m{3.00 cm} !{\vrule width 1.25 pt}
				>{\centering\arraybackslash} m{3.00 cm} !{\vrule width 1.25 pt}
				>{\centering\arraybackslash} m{3.00 cm} 
			}
			\Xhline{4\arrayrulewidth}
			\multicolumn{3}{c}{\textbf{Variables and their values}}
			\\
			\Xhline{\arrayrulewidth}
			Variable
			&
			Company A
			&
			Company B
			\\
			\Xhline{3\arrayrulewidth}
			A & 20 & 90
			\\
			\Xhline{1\arrayrulewidth}
			B & 25 & 60
			\\
			\Xhline{1\arrayrulewidth}
			C & 25 & 90
			\\
			\Xhline{1\arrayrulewidth}
			D & 100 & 100
			\\
			\Xhline{1\arrayrulewidth}
			E & 80 & 10
			\\
			\Xhline{1\arrayrulewidth}
			F & 90 & 15
			\\
			\Xhline{1\arrayrulewidth}
			G & 25 & 25
			\\
			\Xhline{1\arrayrulewidth}
			H & 60 & 60
			\\
			\Xhline{1\arrayrulewidth}
			\Xhline{1\arrayrulewidth}
			I & 15 & 75
			\\
			\Xhline{4\arrayrulewidth}
		\end{tabular}
	\end{tcolorbox}
	\renewcommand{\arraystretch}{1.0}
\end{table}

\begin{figure}[H]
	\begin{tcolorbox}
		[
			grow to left by = - 0.00 cm,
			grow to right by = - 0.00 cm,
			colback = white,					
			enhanced jigsaw,					
			sharp corners,
			toprule = 1.0 pt,
			bottomrule = 1.0 pt,
			leftrule = 0.1 pt,
			rightrule = 0.1 pt,
			sharp corners,
			center title,
			fonttitle = \bfseries
		]
		\centering
		\includegraphics[scale = 0.40, trim = {0 0 0cm 0}, clip]{"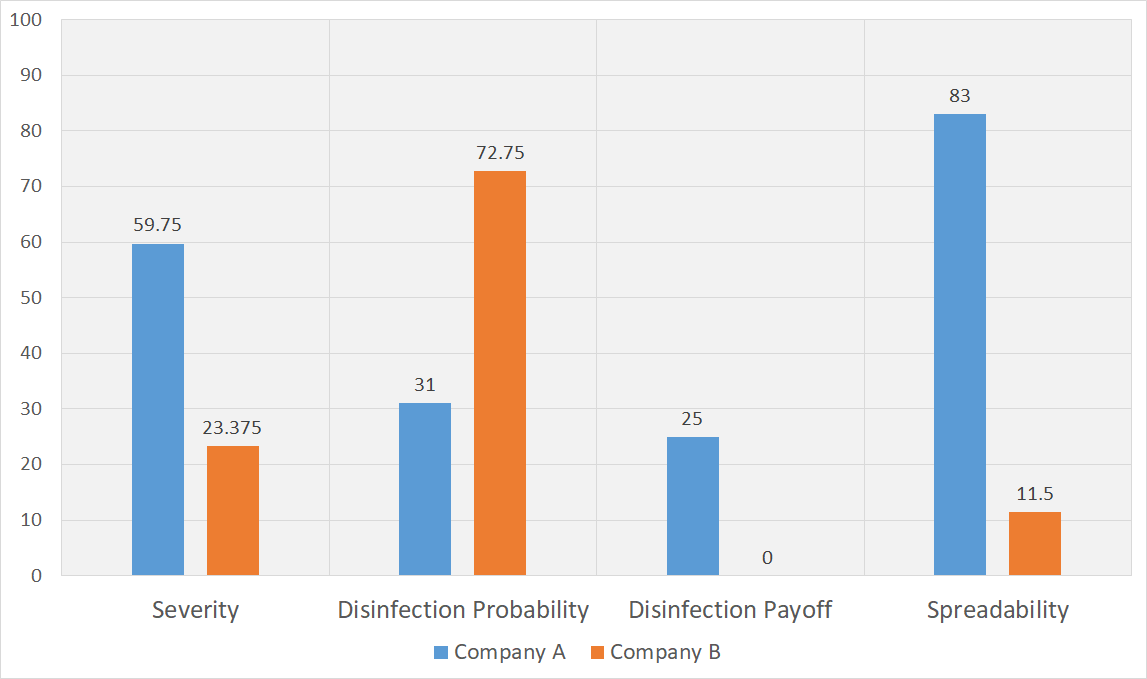"}
		\caption{Graphic representation of the outcome provided by the preceding formulas given the traits that constitute the profile of the $2$ different companies. Its clear that the company A faces a significantly more severe problem compared to company B since its traits as a player are less beneficial than A's. The spreadability also rises significantly in the case of company A since the employees are technologically illiterate and the initial spread makes their condition worse. Looking at the aformentioned formulas and the scenario bellow we can get a better understanding of the outcome in the graph. The calculations were made through a computer program written in C based on the mathematical concepts introduced above.}
		\label{fig:Math Graph 1}
	\end{tcolorbox}
\end{figure}

The above Figure \ref{fig:Math Graph 1} shows the different results the formulas produced after inserting values based of $2$ different fictional companies, one that is technologically literate and one that isn't. Keep in mind that some factors matter less in this scenario with these formulas. In this case a ransomware attack like VirLock’s could be much more effective against them in comparison to a company that keeps their employees up to date on the on-going threats. That is because viruses like VirLock aim for this trait in their victims. When VirLock gets sent via email to the aforementioned employees pretending to be an important document that they should forward, they will probably not think twice before spreading the virus. This will affect the spreadability score of a virus, and, thus, the severity as well. Assuming that the amount of data and their criticality is basically the same in both companies and that both of them are getting attacked from the same virus, we can understand the advantages and disadvantages of those $2$ players against the virus.

In the case of the technologically illiterate company the variables $A, F$ and $I$ have great differences compared to the technologically literate one, since most of the employees wouldn’t be able to identify a candy wrapped scam email and probably forward it to the rest of the network, i.e., the $F$ variable assumes higher values. By the time they realize that an infection took place they wouldn’t have a thought-out strategy of disinfection and the risk of disinfecting operations during that time could have a worse effect. In contrast, a company with well trained and up to date employees would probably have less of a problem and would restrict the spread in their network much faster. The $F$ variable would be lower and variable $I$ higher, leading to a less severe attack as well as lower spreadability and generally a higher payoff for them. In addition to the above the technologically literate company probably has much more data that are of critical importance, but have a much better data to infected data ratio due to their proactive strategy against the attack. In addition to that, their economical state is superior to the other company. The technologically illiterate company has a much higher data to infected data ratio but their data may not be of such critical importance. In such a scenario one may modify the values of the $B, C, D$, and $E$ variables accordingly.

Additionally, we have run extensive experiments in which a single trait was fixed to a constant value. This approach has enabled us to evaluate how a certain variable corresponding to that trait  affects the remaining variables and to what extent. In the following figures, a variable is chosen to be constant while all the other variables take values from $0$ to $100$. Examining the charts can visually illustrate the effects of the traits. Further explanations are given in the corresponding captions of the figures.

\begin{figure}[H]
	\begin{tcolorbox}
		[
			grow to left by = - 0.00 cm,
			grow to right by = - 0.00 cm,
			colback = white,					
			enhanced jigsaw,					
			sharp corners,
			toprule = 1.0 pt,
			bottomrule = 1.0 pt,
			leftrule = 0.1 pt,
			rightrule = 0.1 pt,
			sharp corners,
			center title,
			fonttitle = \bfseries
		]
		\centering
		\includegraphics[scale = 0.40, trim = {0 0 0cm 0}, clip]{"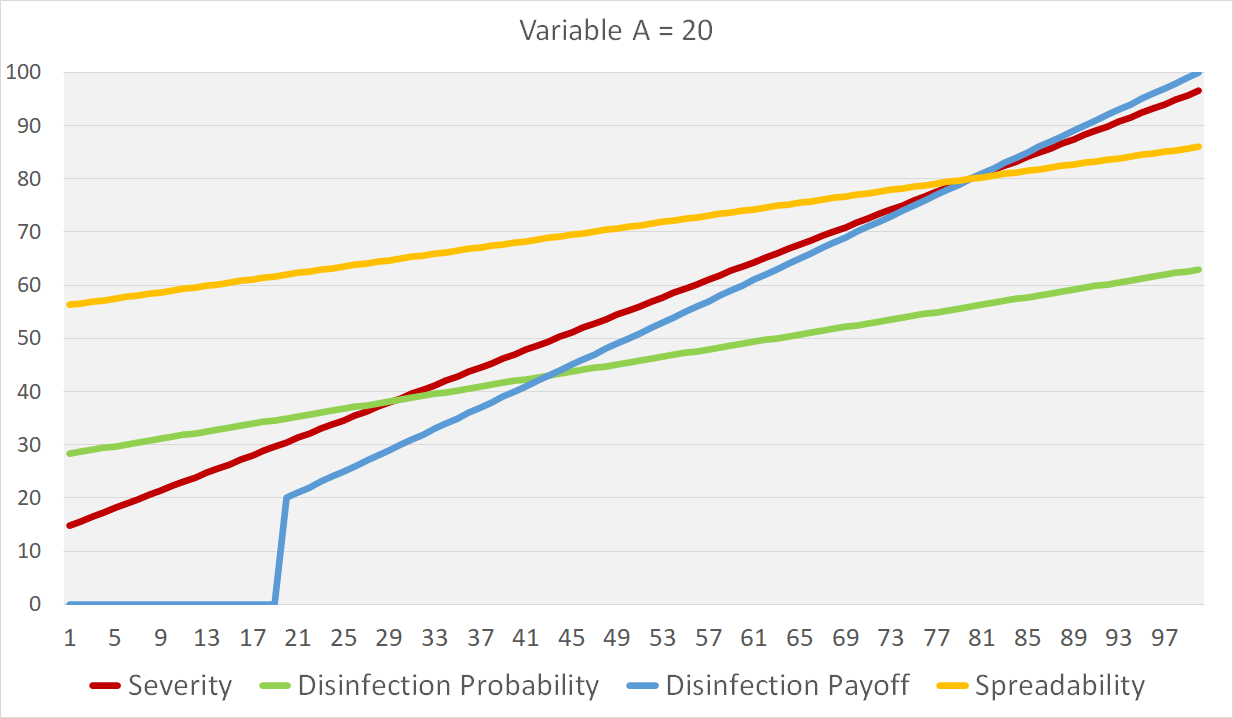"}
		\caption{This chart displays a graphic plot of the formulas in the case where variable $A$ is constant with value $20$, while the values of the rest of the variables range from $0$ to $100$.}
		\label{fig:Math Variable A 20}
	\end{tcolorbox}
\end{figure}

\begin{figure}[H]
	\begin{tcolorbox}
		[
			grow to left by = - 0.00 cm,
			grow to right by = - 0.00 cm,
			colback = white,					
			enhanced jigsaw,					
			sharp corners,
			toprule = 1.0 pt,
			bottomrule = 1.0 pt,
			leftrule = 0.1 pt,
			rightrule = 0.1 pt,
			sharp corners,
			center title,
			fonttitle = \bfseries
		]
		\centering
		\includegraphics[scale = 0.40, trim = {0 0 0cm 0}, clip]{"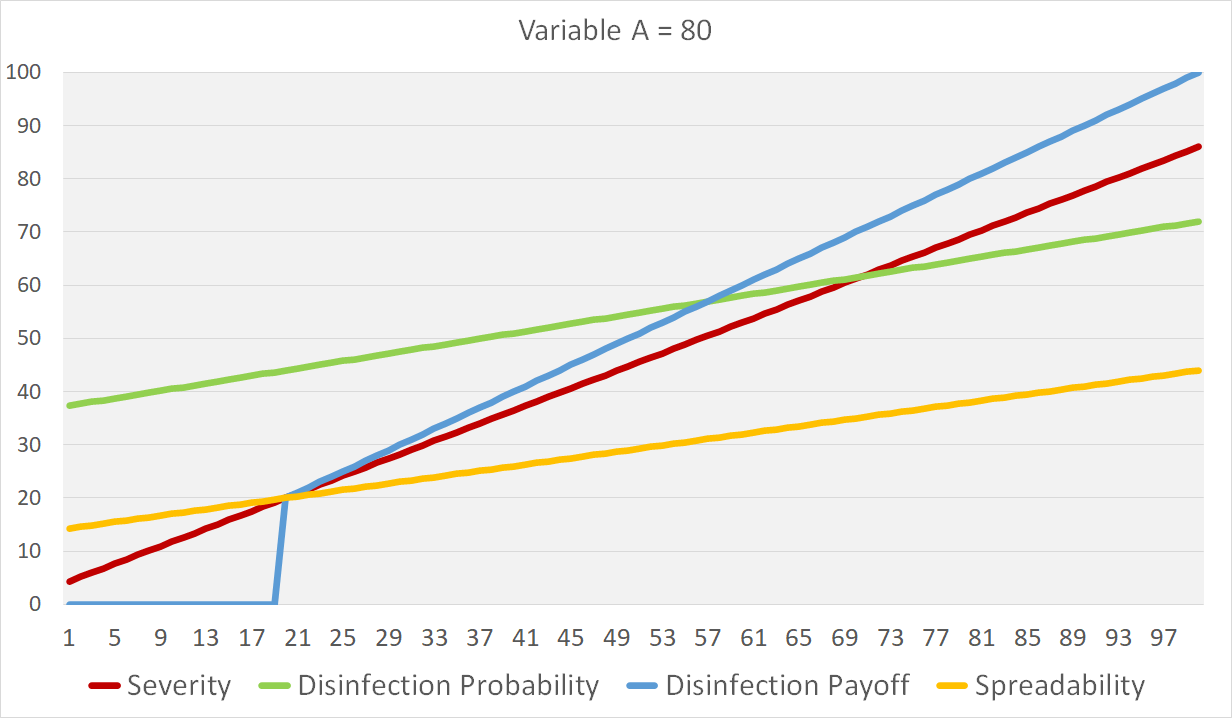"}
		\caption{This chart plots the formulas when the variable $A$ is constant with value $80$ while the values of the rest of the variables range from $0$ to $100$.}
		\label{fig:Math Variable A 80}
	\end{tcolorbox}
\end{figure}

Figure \ref{fig:Math Variable A 20} makes it even more clear that the huge increase in the values of the severity output from the formulas depends heavily on the value of the variable $A$. The conclusion is, as expected, that a company or an individual that is technologically illiterate and not up to date to the virus ``games'' is not only more susceptible to infection, but also finds it harder to disinfect and not spread the virus (especially in the scenario of VirLock) which is also why the spreadability line is much higher than bellow. Note that the severity line is slightly higher but steeper and the disinfection probability is also higher. A simple comparison between Figure \ref{fig:Math Variable A 20} and Figure \ref{fig:Math Variable A 80} demonstrates that the disinfection payoff is the same in both cases, i.e., when $A = 20$ and when $A = 80$, since this formula doesn't make use of the variable $A$.

\begin{figure}[H]
	\begin{tcolorbox}
		[
			grow to left by = - 0.00 cm,
			grow to right by = - 0.00 cm,
			colback = white,					
			enhanced jigsaw,					
			sharp corners,
			toprule = 1.0 pt,
			bottomrule = 1.0 pt,
			leftrule = 0.1 pt,
			rightrule = 0.1 pt,
			sharp corners,
			center title,
			fonttitle = \bfseries
		]
		\centering
		\includegraphics[scale = 0.40, trim = {0 0 0cm 0}, clip]{"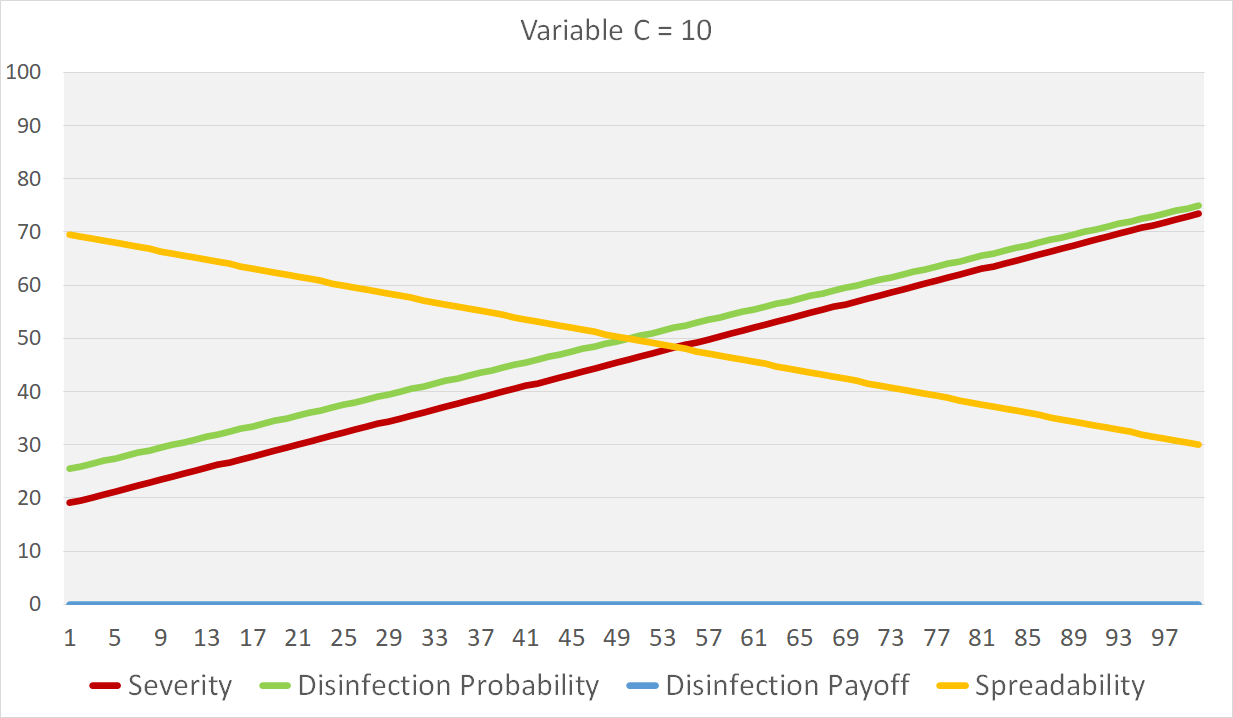"}
		\caption{The above figure contains a graphic plot of the formulas in the case where variable $C$ is constant with value $10$ while the values from the rest of the variables range from $0$ to $100$.}
		\label{fig:Math Variable C 10}
	\end{tcolorbox}
\end{figure}

\begin{figure}[H]
	\begin{tcolorbox}
		[
			grow to left by = - 0.00 cm,
			grow to right by = - 0.00 cm,
			colback = white,					
			enhanced jigsaw,					
			sharp corners,
			toprule = 1.0 pt,
			bottomrule = 1.0 pt,
			leftrule = 0.1 pt,
			rightrule = 0.1 pt,
			sharp corners,
			center title,
			fonttitle = \bfseries
		]
		\centering
		\includegraphics[scale = 0.40, trim = {0 0 0cm 0}, clip]{"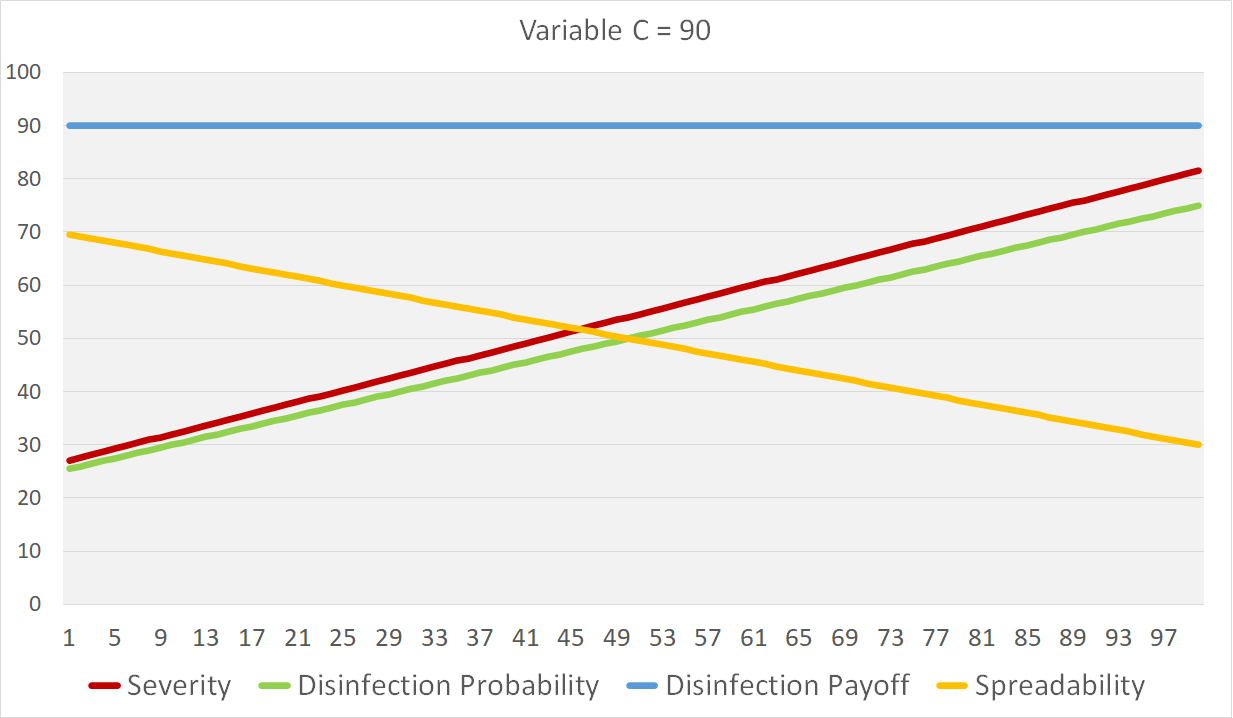"}
		\caption{Chart presenting the output of the formulas where the variable $C$ is constant with value $90$ while the values from the rest of the variables range from $0$ to $100$.}
		\label{fig:Math Variable C 90}
	\end{tcolorbox}
\end{figure}

In both Figures \ref{fig:Math Variable C 10} and Figures \ref{fig:Math Variable C 90}, we may observe a cross shape between severity and disinfection payoff in both cases where $C = 10$ and $C = 90$ chart. The cross is obviously higher in the $C = 90$ chart since the criticality of the data is extremely high, something that makes the attack more severe. The main difference between the two charts is the disinfection payoff of employing any strategy to get the data back. The intuition behind this is that since the criticality of the data encrypted is that high, it is extremely important (with value proportionate to the criticality of the data) to employ any strategy to get them back. The exact opposite from the chart where $C = 10$ where the $DP = 0$ since its not worth it to employ anything with such a low level data criticality.

\section{Discussion and conclusions} \label{sec:Discussion and Conclusions}

This article broadens the outlook on a potential new perspective and offers a starting point for experimentation in the future. Through the correlation of certain computer viruses with similar biological ones we set ground for further research involving the game theoretic perspective. Studying the traits and their similarities, apart from better understanding their impact factor in a game, can also hint that remedies and disinfection mechanics that work well in the biological world could appear effective through the proper correlation to the computer world in certain cases.

Game theory is a tool that might prove useful in the background as behavioral analysis is more and more needed with most antimalware software constantly improving their analysis tools. Analyzing the behavior of viruses, as well as the host machines and the users actions', can lead to the formulation of more effective strategies against an opponent. Using game theory as a tool could help tackle never-seen-before threats assisting AI technology. Alternative strategies for dealing with biological and computer viruses can appear through study of the relationship between them using appropriate tools and methodology. These game theory tools and methods we discussed could also prove useful in enhancing penetration testing reports, and vise versa, since the results of the penetration testing could feed the values of variables used in games.

The topic of threat assessment is very important in practice when trying to tackle multiple scenarios, especially in environments where data is critical. Moreover, risk analysis, in the form attempted in the previous tables and charts by analyzing the disinfection steps, is equally important in order to select a strategy that will offer the ultimate payoff. There are factors that sometimes are not taken into account, even though they are important and could change the possible outcome of an attack. The specific users and their traits should be taken into account when picking a strategy in order to achieve the optimum outcome, since just picking the ultimate security doesn't factor costs like price, complexity, application etc.

As a future work we intend to experiment with lab tests on computer viruses' spread in medium and small-sized networks. In that respect, the introduction of new games and pay-off matrices, based on real life scenarios for a wider range of users will be instrumental. We believe that evolutionary game theory provides the appropriate tools for keeping better track of viruses' mutations and describing strategies for fending off infections. To this end, we plan to extend the current formulas and further evaluate their usefulness for disinfection, spreadability and severity. Hopefully, this will allow us to obtain even more objective and realistic measurements from the corresponding games regarding the severity of a computer virus of a certain type in certain scenarios, the probability of disinfection, as well as if some specific strategies are more effective for this type of virus. Furthermore we would like to verify whether processes like natural selection and its properties also apply to computer environments. 

It's important to point out that such formulas and calculations require a significant amount of data from reputable sources and security experts, as well as systematic maintenance to keep them up to date. It is clear that our formulas are still in an early form and they cannot apply to every possible scenario just yet, but they are an important tool in a theoretically constructed game theoretic framework for evaluation and assessment of effective strategies, assisting further research. By compiling and analyzing a great amount of data from real networks, small and big, company and personal, secured or not, that suffered virus attacks the accuracy of the parameters and constants involved could be increased, while new factors and variables could appear that should be taken into account and are not yet. This in turn will provide the key to evolve the formulas so that they better reflect real-world situations. 

Another avenue of possible future research could be the assessment of already known and used biological virus formulas that estimate corresponding viral factors as well as their spreadability and severity per population. Again, by correlating technology and biological data interesting results could appear (for example the population of a city could be correlated to a company’s computer network). The above can show us the importance of certain traits and help us think of better pay-offs in the pay-off matrices of appropriately constructed games.

\bibliographystyle{ieeetr}
\bibliography{ViewingBiological&ComputerVirusesManifestationsOfGames}

\begin{thebibliography}{10}
\bibitem{uniserve}
``What are the different types of computer viruses?.''
\bibitem{Norton2019}
NortonLifeLock, ``What is a computer worm, and how does it work?.''
\end{thebibliography}

\end{document}